\documentclass[preprintnumbers,nofootinbib,preprint,12pt,showpacs]{revtex4}

\usepackage{bm,amsmath,amssymb}

\usepackage[dvips]{graphicx}

\allowdisplaybreaks[4]

\begin{document}

{~}

\vspace{1cm}

\title{\Large 
Geodetic Precession \\ in Squashed Kaluza-Klein Black Hole Spacetimes
\vspace{1cm}
}

\author{
Ken Matsuno\footnote{E-mail: matsuno@sci.osaka-cu.ac.jp} and  
Hideki Ishihara\footnote{E-mail: ishihara@sci.osaka-cu.ac.jp}
}	

\address{
  Department of Mathematics and Physics, Graduate School of Science, 
  Osaka City University, 
  3-3-138 Sugimoto, Sumiyoshi-ku, Osaka 558-8585, Japan
\vspace{2cm}
}

\begin{abstract}

We investigate the geodetic precession effect of a parallely transported 
spin-vector along a circular geodesic 
in the five-dimensional squashed Kaluza-Klein black hole spacetime. 
Then we derive the higher-dimensional correction of the precession angle to 
the general relativity. 
We find that the correction is proportional to the square of (size of
extra dimension)/(gravitational radius of central object).

\end{abstract}

\preprint{OCU-PHYS 319}  
\preprint{AP-GR 71}

\pacs{04.50.-h,~ 04.70.Bw}

\date{\today}

\maketitle

\section{Introduction}

Recently, classical general relativity in higher dimensions, 
suggested by superstring theory, has gathered much attention. 
Especially, many studies are devoted to higher-dimensional 
black holes (see \cite{EmparanReal} for review) 
because black holes would be expected as windows to extra dimensions. 
One of the most interesting problems is   
a verification of extra dimensions by physical phenomena related to 
the higher-dimensional black holes.
Higher-dimensional black holes in accelerators \cite{Banks:1999gd, Dimopoulos:2001, Giddings:2001bu} 
and in cosmic ray \cite{ArkaniHamed:1998nn, Argyres:1998qn, Feng:2001ib, Anchordoqui:2001cg}   
and gravitational waves from higher-dimensional black holes \cite{Seahra:2004fg}
are studied for this purpose.

In astrophysics, 
the study of the classical tests of the general relativity: 
the light deflection, the perihelion shift, the time delay, 
could provide methods
to know properties of the spacetime metrics around compact objects. 
In addition, the precession of a gyroscope \cite{GeoE, FDE, GPB} is another 
experimental way to reveal the spacetime structure.
The studies of these effects open the possibility of testing higher-dimensional 
models by using 
astronomical and astrophysical observations at the solar system scale.
Indeed, such problems are considered by using 
five-dimensional two-parameter family of Kaluza-Klein type solutions with 
an S$^2$ symmetry \cite{LimW,KWE,LOW,LW2,Over,LO}, 
and S$^2$ symmetric solutions in brane world models \cite{LW,BHL,JMS,Seahra:2002xe}.

Note that the exterior spacetime is 
described by the Schwarzschild metric 
for standard general relativistic spherical compact objects. 
However, in higher-dimensional spacetime models, the exterior
metric of a static star is no longer the Schwarzschild metric. 
In this paper, we consider another class of Kaluza-Klein type metrics 
which would describe exterior of compact objects. 
In the higher-dimensional spacetimes, even if   
we impose the asymptotic flatness to the four-dimensional part of the spacetime,    
there are various possibilities of fiber bundle structures of 
the extra dimensions as the fiber over the four-dimensional base spacetime.  
The black hole solutions with non-trivial bundle structures 
have been studied by various authors.  
Dobiasch and Maison found the first 
five-dimensional vacuum Kaluza-Klein black hole solution    
which asymptotes to a twisted S$^1$ fiber bundle over the four-dimensional 
Minkowski spacetime \cite{DM2}, 
and Gibbons and Wiltshire clarified the meaning of the metric \cite{GW}.
Later, the solution was generalized to the rotating case \cite{Rash2}, and  
charged case in the five-dimensional Einstein-Maxwell theory \cite{IM}. 
It is shown that this family of black holes have squashed S$^3$ horizons,  
and that the metrics behave as fully five-dimensional black holes in the vicinity 
of horizon, while behave as four-dimensional black holes in the far region. 
The squashed Kaluza-Klein black hole solutions of cohomogeneity-one 
are related to asymptotically flat black hole solutions 
by a squashing function \cite{IM, TW}. 
Some of the supersymmetric black hole solutions \cite{GSYhole,IKMTmulti,MINT} 
in the framework of ref. \cite{GGHPR}  
and non-supersymmetric solutions \cite{NIMT,TIMN,TI,SSW,TYM,GS} were constructed 
by the procedure of squashing.
These solutions also have been generalized in various directions, e.g., 
in Einstein-Yang-Mills theory \cite{BR}, and 
in Einstein-Maxwell-dilaton theory \cite{SSY,Allahverdizadeh:2009ay}.

Motivated by the Gravity Probe B experiment \cite{GPB}, 
we focus our attention to the precession of a gyroscope in an orbit around  
the Earth. We assume that the geometry of the region outside the Earth is described by 
the five-dimensional vacuum static squashed 
Kaluza-Klein black hole metric \cite{DM2, IM}. 
We derive the higher-dimensional correction of the precession angle to 
the general relativity, which is related with the size of the extra dimension.

We regard a rotating axis of the gyroscope carried by a satellite 
as a spacelike spin-vector. 
While the satellite moves along the stable circular orbit around the Earth,  
the spin-vector is parallely transported along this timelike circular geodesic. 
When the satellite returns to the initial position, 
an angular difference of the final spin-vector from the initial one appears. 
This is the geodetic effect \cite{GeoE}. 
Since the Earth rotates, the spin-vector is dragged around the rotation axis 
of the Earth by the Kerr geometry \cite{FDE}. 
The frame dragging effect by the Kerr rotation is quite smaller than  
the geodetic effect, 
then, we investigate the correction by the extra dimensions 
to the geodetic effect in this paper.

This paper is organized as follows. 
In the next section, 
we review the properties of five-dimensional static squashed Kaluza-Klein black 
hole solution.   
In section \ref{geodesicsec}, 
we show that the five-dimensional squashed Kaluza-Klein black hole spacetime admits
stable circular orbits similar to the four-dimensional Schwarzschild black holes.
In section \ref{geodeticsec}, 
we consider the parallel transportation of a spin-vector 
along a circular orbit 
in the squashed Kaluza-Klein black hole spacetime, 
then derive the geodetic precession angle with the higher-dimensional correction. 
Finally, we devote section \ref{conclsec} to discussion.

\section{Review of Squashed Kaluza-Klein Black Holes}

We start from the five-dimensional vacuum Einstein equation,  
\begin{equation}
	R_{\mu\nu} = 0.                  
\label{EinEq}
\end{equation}
The static Kaluza-Klein black hole with a squashed S$^3$ horizon 
 \cite{DM2, GW} is one of the exact solutions of \eqref{EinEq}.  
The metric is written as \cite{IM}   
\begin{align}
 &ds^2 = - V(\rho ) dt ^2 + U(\rho ) 
\left [ \frac{d\rho ^2}{V(\rho )} + \rho ^2 d \Omega_{S^2} ^2 \right]
 + \frac{r_\infty ^2}{4 U(\rho )}\left( d \psi + \cos \theta d \phi \right) ^2 , 
\label{BHmet}
\\
&d \Omega_{S^2} ^2 := d \theta ^2 + \sin ^2 \theta d \phi^2, 
\end{align}
where the functions $V(\rho ),~U(\rho )$ are given by
\begin{align}        
 V(\rho) =  1- \frac{\rho _g}{\rho}, \quad 
 U(\rho) = 1 + \frac{\rho_0}{\rho}, 
\end{align}
and positive parameters $r_\infty, ~ \rho _g$ and $ \rho _0$ are related 
as $ r_\infty^2 = 4 \rho_0 \left( \rho _g + \rho_0 \right)$. 
The ranges of coordinates are 
$-\infty <t< \infty ,~0<\rho <\infty ,~0\leq \theta \leq \pi ,~
0\leq \phi \leq 2\pi $, and $0\leq \psi \leq 4\pi $.

The black hole horizon is located at $\rho = \rho _g $, 
and the curvature singularity at $\rho = 0$ is concealed behind it. 
The induced metric on the three-dimensional cross section of 
the black hole horizon with a time slice $t= const$. is obtained as 
\begin{align}        
 \left. ds^2 \right|_{\rho = \rho _g, ~ t = const.} = 
 \rho _g \rho _0 \left[ \left( 1 + \frac{\rho _g}{\rho _0} \right) d \Omega_{S^2} ^2 
+ \left( d \psi + \cos \theta d \phi \right) ^2  \right] .   
\label{IndMet}
\end{align}
Then, we see that the metric of 
black hole horizon 
is the squashed S$^3$ in the form of the Hopf bundle. 
Since $\rho _g > 0$ and $\rho _0 > 0$, 
we see that the radius of the S$^2$ base is larger than that of the S$^1$ fiber.

At the infinity, $\rho = \infty$, 
the metric \eqref{BHmet} behaves as 
\begin{align}
 ds^2 \simeq - dt ^2 + d\rho ^2 + \rho ^2 d \Omega_{S^2} ^2 
 + \frac{r_\infty ^2}{4}  
 \left( d \psi + \cos \theta d \phi \right) ^2  .
\end{align}
That is, the metric \eqref{BHmet} asymptotes to a twisted constant S$^1$ fiber 
bundle over the four-dimensional Minkowski spacetime. 
The size of compactified extra dimension of the spacetime \eqref{BHmet} 
at the infinity is given by $r_\infty$. 
The Komar mass\footnote{ For the squashed black hole, the Komar mass 
takes a different value of the Abbott-Deser mass \cite{Kurita:2007hu, Kurita:2008mj}.
}
of this black hole \eqref{BHmet} is given by  
\begin{gather}
 	M =\frac{\pi r_\infty \rho _g}{G_5} 
		=\frac{\rho _g}{2 G_4} ,
\label{massofbh}
\end{gather}
where 
the five-dimensional gravitational constant $G_5$ and the four-dimensional 
one $G_4$ are related as
\begin{align} 
	G_4 = \frac{G_5}{2\pi r_\infty} .  
\end{align}

Here, 
we discuss the physical meanings of the parameter $\rho _0$ in \eqref{BHmet}. 
If $\rho _0 \ll \rho _g$, 
$\rho$ dependence of the function $V(\rho )$ is important but the function 
$U(\rho )$ is almost unity for an observer outside the horizon, 
$\rho _0 \ll \rho _g \lesssim \rho$. 
Then, the observer feels the spacetime 
as the four-dimensional Schwarzschild  
black hole with a constant twisted S$^1$ fiber with the metric
\begin{align}
 ds^2 \simeq - \left( 1 - \frac{\rho _g}{\rho} \right) dt ^2 
	+ \left( 1 - \frac{\rho _g}{\rho} \right) ^{-1} d\rho ^2 
	+ \rho ^2 d \Omega_{S^2} ^2 
        + \frac{r_\infty^2}{4} \left( d \psi + \cos \theta d \phi \right) ^2 .
\end{align}
On the other hand, if $\rho _g \ll \rho _0$, 
the function $U(\rho )$ becomes important 
for an observer at $\rho _g \lesssim \rho \ll \rho _0$.
With the help of a new coordinate $r = 2 \sqrt{\rho_0 \rho }$ and 
a parameter $r_g = 2 \sqrt{\rho_0 \rho _g }$, 
since $r_g^2 \ll r_\infty^2 $, 
the metric \eqref{BHmet} approaches to 
the five-dimensional Schwarzschild-Tangherlini black hole \cite{Tang}: 
\begin{align}
 ds^2 \simeq  ds^2_\text{ST} 
	= - \left( 1 - \frac{r_g ^2}{r^2} \right) dt^2 
       + \left( 1- \frac{r_g ^2}{r^2} \right)^{-1} dr^2 
       + r^2 d \Omega_{S^3} ^2,                                    
\label{SchBHmet2}
\end{align}
where $d \Omega_{S^3}^2$ denotes the metric of the unit three-sphere.  
Then, the observer feels the spacetime 
as an almost $S^3$ symmetric black hole. 
Therefore, the parameter $\rho _0$ gives the typical scale of transition 
from five-dimension to effective four-dimension.

\section{Circular Orbits around Squashed Kaluza-Klein Black Holes}
\label{geodesicsec}

We consider timelike geodesics in the five-dimensional squashed Kaluza-Klein 
black hole.  
The Lagrangian for a test particle in the metric \eqref{BHmet} is   
\begin{align}
 \mathcal L = \frac{1}{2} \left[ 
- V \dot t^2 + \frac{U}{V} \dot \rho ^2 + U\rho ^2 \left( \dot \theta ^2 + \sin^2 \theta \dot \phi ^2 \right) 
+ \frac{r_\infty ^2}{4U} \left( \dot \psi + \cos \theta \dot \phi \right) ^2
\right] ,
\end{align}
where the overdot denotes the differentiation with respect to   
the proper time $\tau$, then we set the condition $2 \mathcal L = -1$. 
We can obtain three constants of motion, 
\begin{align}
 	E &:= V \dot t , 
\\
 	L &:= U\rho ^2 \sin^2 \theta \dot \phi 
		+ \frac{r_\infty ^2 \cos\theta}{4U} 
		\left( \dot \psi + \cos \theta \dot \phi \right) , 
\\
 	p_\psi &:= \frac{r_\infty^2}{4U} \left(\dot\psi+\cos\theta\dot\phi\right) .
\end{align}

\begin{figure}[htbp]
 \begin{center}
 \includegraphics[width=15cm,clip]{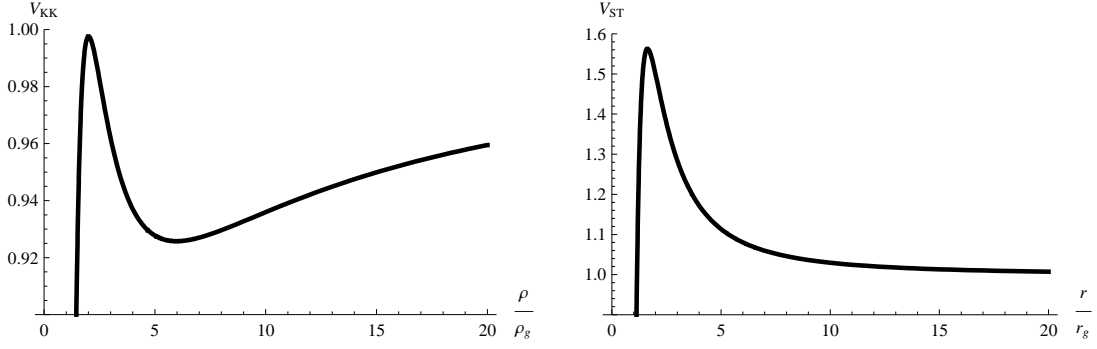}
 \end{center}
 \caption{
Effective potentials for a massive test particle moving 
in the five-dimensional squashed Kaluza-Klein black hole metric (left). 
We set the parameters as $L = 2 \rho _g$ and $\rho _0 = 10^{-2} \rho _g$ 
in \eqref{effKK}. 
The same one in the five-dimensional Schwarzschild-Tangherlini black hole 
metric (right) with the parameters $L_\text{ST} = 2 r_g$  in  \eqref{effSch}. 
}
 \label{pots}
\end{figure}

Here, we assume that the particle has no momentum in the extra direction,
i.e., $p_\psi = 0$. 
The effective Lagrangian for the particle is 
\begin{align}
 	\mathcal L _\text{eff} 
		= \frac{1}{2} \left[
		- V \dot t^2 + \frac{U}{V} \dot \rho ^2 
		+ U\rho ^2 \left( \dot \theta ^2 + \sin^2 \theta \dot \phi ^2 \right) 
		\right] ,
\end{align}
where $2 \mathcal L _\text{eff} = -1$. 
We see that this effective Lagrangian has the same form in the case of 
four-dimensional spherically symmetric spacetimes. 
Then, we can concentrate on orbits with $\theta = \pi/2$   
on the assumption of $p_\psi = 0$.   
In these conditions, we obtain the energy conservation equation
\begin{align}
  \left( 1 + \frac{\rho_0}{\rho} \right)\left( \frac{d \rho}{d \tau } \right)^2 
 	+ V_\text{KK}(\rho) = E^2 , 
\label{geodesicKK} 
\end{align}
where the effective potential is given by
\begin{align}
 V_\text{KK} (\rho ) = \left( 1- \frac{\rho_g}{\rho} \right) 
                \left( 1 + \frac{L ^2}{\rho \left( \rho + \rho_0 \right)} \right).  
\label{effKK}
\end{align}
We note that  \eqref{geodesicKK} with \eqref{effKK} 
reduces to the energy conservation equation 
of a particle in the four-dimensional Schwarzschild black hole spacetime 
in the limit $\rho _0 \to 0$. 
A typical shape of the effective potential $V_\text{KK}$ 
is shown in the left panel of Figure \ref{pots}. 
We see that there is a stable circular orbit at the local minimum of the effective potential.

The squashed Kaluza-Klein black holes, which have stable circular orbits,  
make remarkable contrast with the higher-dimensional asymptotically flat 
black holes, which have no stable bound state of particles.  
For the simplest example, 
we consider timelike geodesics in 
the five-dimensional Schwarzschild-Tangherlini black hole spacetime 
with the metric \eqref{SchBHmet2}. 
We consider motions in an equatorial plane, that is, $\theta = \pi /2$, then 
we have the energy conservation equation 
\begin{align}
 \left( \frac{dr}{d \tau} \right)^2 + V_\text{ST}(r) &= E_\text{ST}^2 , 
\end{align}
with the effective potential 
\begin{align}
 V_\text{ST} (r) &= \left( 1-  \frac{r_g ^2}{r^2}  \right) 
              \left( 1 +  \frac{L_\text{ST} ^2}{r^2} \right),     
\label{effSch}
\end{align}
where $E_\text{ST}$ and $L_\text{ST}$ are constants of motion of the system \cite{FS}. 
A typical shape of the effective potential 
is drawn in the right panel of Figure \ref{pots}.  
Clearly, there is no stable circular orbit. 
The metric of squashed Kaluza-Klein black hole \eqref{BHmet} 
with a condition $\rho _0 \ll \rho _g$ 
can be a candidate for the metric which describes a spacetime around the Earth 
where a test particle can travel along a circular orbit. 
We can expect the appearance of the higher-dimensional correction, 
which is related to the parameter $\rho_0$, to 
the geodetic precession effect of four-dimensional relativity.

In the present paper, we restrict ourselves to a circular motion $\rho = R = \text{const.}$ 
with $p_\psi = 0$ and $\theta = \pi / 2$. 
Then we have 
\begin{align}
 	u^\rho =0, \quad u^{\theta}=0, \quad \text{and}\quad
	u^\psi=0,  
\end{align} 
where $u^\mu = dx^\mu /d\tau $ is the five-velocity of the particle. From 
$V_\text{KK} = E^2$ and $dV_\text{KK}/d\rho = 0$, 
we have 
\begin{align}
	E^2 =  \frac{(R-\rho_g)^2(2R+\rho_0)}{R^2(2R-3\rho_g) + R\rho_0(R-2\rho_g)} , 
\quad
	L^2 = \frac{\rho_g R(R+\rho_0)^2}{R(2R-3\rho_g )+ \rho_0(R-2\rho_g)}. 
\end{align}
and 
\begin{align}
 u^t & 
	= \sqrt{ \frac{R(2R+\rho_0 )}{R(2R-3\rho_g)+\rho_0(R-2\rho_g)} } ,
\label{ut}
\\
 u^\phi & 
	= \sqrt{\frac{\rho_g}{R^2 (2R-3\rho_g )+\rho_0 R(R-2\rho_g )} } .
\label{uphi}
\end{align}

By using \eqref{massofbh}, \eqref{ut} and \eqref{uphi},
we obtain Kepler's third law 
in the squashed Kaluza-Klein black hole spacetime 
as  
\begin{align}
 T^2 
	= \frac{4\pi^2}{G_4 M} R^3
	\left(1 
		+\frac{1}{2}\frac{\rho_0}{R}\right) ,
\end{align}
where $T$ 
denotes the orbital period. 
The second term in the right hand side is the correction by the extra dimension.

\section{Verification of Extra Dimension 
by Geodetic Effect Observation}
\label{geodeticsec}

We regard 
the rotating axis of the gyroscope carried 
by a satellite as a spacelike spin-vector $S^\mu$ 
parallely transported along a timelike geodesic with the five-velocity $u^\mu$. 
The parallel transporting equation of $S^\mu $ in the direction 
of $u^\mu $ is
\begin{align}
	u^{\mu } \nabla _{\mu} S^{\nu }= 0 .
\label{eqofparatrans}
\end{align}
We impose the orthogonality condition between $u^\mu $ and $S^\mu $,
and normalization condition
\begin{align}
	u^{\mu } S _{\mu} = 0,   \quad S^{\mu } S_{\mu }= 1. 
\label{eqoforthogonal}
\end{align}

The parallel transporting equation \eqref{eqofparatrans} along the circular orbits 
in the squashed Kaluza-Klein spacetime becomes
\begin{align}
 & \frac{dS^t}{d\tau } + \frac{\rho _g u^t}{2R(R-\rho _g)} S^\rho = 0 , \\
 & \frac{dS^\rho }{d\tau } + \frac{\rho _g (R-\rho _g) u^t}{2R^2(R+\rho _0)} S^t 
   - \frac{(R-\rho _g) (2R +\rho _0) u^\phi }{2(R+\rho _0)} S^\phi  = 0 , \\
 & \frac{dS^\phi }{d\tau } + \frac{(2R + \rho _0) u^\phi }{2R(R+\rho _0)} S^\rho = 0 , \\   
 & \frac{dS^\theta }{d\tau } + \frac{\rho _0 (\rho _g +\rho _0) u^\phi}{2(R+\rho _0)^2} S^\psi = 0 , \\  
 & \frac{dS^\psi }{d\tau } - \frac{u^\phi}{2} S^\theta  = 0 .
\end{align}
By using \eqref{ut} and \eqref{uphi} for the explicit forms of $u^t$ and $u^\phi$,  
we obtain the spin-vector $S^\mu $ as  
\begin{align}
 S^{t } &= C^t \sin \left( \Omega \tau \right), \\
 S^{\rho } &= C^\rho \cos \left( \Omega \tau \right), \\
 S^{\phi } &= \frac{C^\phi }{R}\sin \left( \Omega \tau \right), \\
 S^{\theta } &= \frac{C^\theta }{R}\cos \left( \tilde \Omega \tau \right), \\ 
 S^{\psi } &= \frac{C^\psi }{R}\sin \left( \tilde \Omega \tau \right) ,      \label{spin}
\end{align}
where $\Omega $ and $\tilde \Omega $ are given as 
\begin{align}      
 \Omega &= \frac{\sqrt{\rho _g (2R + \rho_0) }}{2 R (R + \rho_0)} , \label{omega} \\
 \tilde \Omega &= \frac{1}{2(R + \rho_0)} \sqrt{\frac{\rho_g\rho_0(\rho_g + \rho_0 )}
{R \left[ R \left( 2R - 3\rho_g \right) + \rho_0 \left( R-2\rho_g\right)\right]} } .
\end{align}
We have imposed initial condition $S^t=S^\phi=S^\psi=0$ at $\tau=0$, 
for convenience.
Constants $C^\mu $ are determined by the orthogonality and 
the normalization condition \eqref{eqoforthogonal} as  
\begin{align}
 C^t &= C^\rho \frac{R + \rho _0}{R - \rho _g} 
\sqrt{ \frac{\rho _g R }{ R \left( 2R - 3\rho _g \right) + \rho _0 \left( R - 2\rho _g \right) } } , 
\\
 C^\phi &= C^\rho \sqrt{ \frac{ R \left( 2R + \rho _0 \right)}
{ R \left( 2R - 3\rho _g \right) + \rho _0 \left( R - 2\rho _g \right) } } , 
\\
 C^\theta  &= \sqrt{\frac{R}{R + \rho_0}
		\left(1-\frac{R+\rho_0}{R-\rho_g} ({C^\rho})^2\right) }, 
\\  
 C^\psi &= \sqrt{\frac{R(R+\rho _0)}{\rho _0 (\rho _g +\rho _0)} 
\left( 1 - \frac{R+\rho _0}{R-\rho _g} {(C^\rho)}^2 \right)} .
\end{align}
Since we consider the situation such as $\rho _0 \ll \rho _g \ll R$,   
we have $\tilde \Omega \ll \Omega$. 
Then, 
we can regard $S^\theta $ and $S^\psi $ are almost constant for some orbital periods. 
Further, we have $C^t \ll C^\phi \simeq C^\rho$.

During one orbital period, $\phi $ goes from $0$ to $2\pi $ and   
a proper time $\tau $ goes from $0$ to $\tau_p := 2 \pi/u^\phi $,  
the geodetic precession angle $\Delta \Theta$ 
is given as
\begin{align}
 \Delta \Theta 
	= \frac{R S^\phi(\tau_p)}{S^\rho(0)} 
	= \left | \Omega \tau_p - 2\pi\right | 
	= \left | 2 \pi \left( \frac{\Omega }{u^\phi } - 1 \right) \right |.  
\label{kakudozure}
\end{align}
Substituting \eqref{uphi} and \eqref{omega} into \eqref{kakudozure},
we obtain the geodetic precession angle in a weak-field limit as 
\begin{align}
	\Delta \Theta 
	= \Delta \Theta _\text{4D} \left( 1+\delta \right) + O \left( R^{-2} \right), 
\end{align}
where 
\begin{align}
 	\Delta \Theta_{\text{4D}} = \frac{3 \pi G_4 M}{R}  
\label{predvalue}
\end{align}
is the predicted value by the four-dimensional Einstein theory and 
\begin{align}
 	\delta = \frac{\rho_0(\rho_0+\rho_g)}{6(G_4 M)^2}
		= \frac{1}{6}\left( \frac{r_\infty}{2G_4 M} \right)^2
\end{align}
is the higher-dimensional correction.  
For $r_\infty \to 0$, equivalently $\rho _0 \to 0$,  
the correction $\delta $ vanishes as expected.   
If the size of extra dimension is $r_\infty \simeq 0.1$ mm, 
we can estimate that $\delta \simeq 10^{-5}$ for the Earth.

\section{Discussions}
\label{conclsec}

In this paper, 
we have considered the precession of a gyroscope in a circular 
orbit around a spherical compact object. 
We have assumed that the five-dimensional squashed Kaluza-Klein black hole solution 
describes the geometry around the compact object. 
We have solved the parallel transporting equation of a spin-vector 
along the circular timelike geodesic 
in the squashed Kaluza-Klein black hole spacetime, 
and derived the geodetic precession angle with the higher-dimensional correction. 
We have shown that the correction to the general relativity 
is in proportion to the square of 
(size of extra dimension)/(gravitational radius of central object). 
If the size of extra dimension is order of 0.1 mm, the correction is 
order of $10^{-5}$ for the Earth.

In the four-dimensional general relativity, the gravitational field 
in vacuum with spherical symmetry is uniquely described by the Schwarzschild 
metric. However, in the higher-dimensional spacetime with Kaluza-Klein 
structure, even if we impose an asymptotic flatness in four-dimensional section, 
the metric is not determined uniquely. 
For example, the direct product of four-dimensional Schwarzschild spacetime 
with a small S$^1$ 
is a possible metric to describe the geometry around the Earth. 
In this case, 
no higher-dimensional correction appears without a momentum of gyroscope in the extra direction. 
In contrast, it is interesting that 
the correction exists even if the gyroscope moves along four-dimensional spacetime 
in the squashed Kaluza-Klein geometry. 
If a precise experiment of gyroscope precession in the Earth orbit agrees 
the expected value of general relativity, 
it requires a rigorous upper limit of the size of extra dimension, or it excludes    
the squashed Kaluza-Klein metric for describing the geometry around the Earth.

\section*{Acknowledgments}
We would like to thank M. Kimura, T. Nakagawa, K. Nakao 
and S. Tomizawa for useful discussions.  
This work is supported by the Grant-in-Aid for Scientific Research No.19540305. 

\newpage


\end{document}